\documentclass[prb,twocolumn,amsmath,amssymb,showpacs]{revtex4}
\usepackage{graphicx}

\def\beq{\begin{eqnarray}}
\def\eeq{\end{eqnarray}}
\def\beqa{\begin{eqnarray}}
\def\eeqa{\end{eqnarray}}

\begin{document}

\title{Charge order and phonon renormalizations:
Possible implications for cobaltates.
}

\author{M. Bejas, A. Greco and A. Foussats}
\affiliation{
Facultad de Ciencias Exactas, Ingenier\'{\i}a y Agrimensura and
Instituto de F\'{\i}sica Rosario
(UNR-CONICET).
Av. Pellegrini 250-2000 Rosario-Argentina.
}

\date{\today}

\begin{abstract}

Several experimental and theoretical studies in cobaltates
suggest the proximity of the system to charge ordering (CO).
We show, qualitatively, 
in the frame of a $t-V$ model coupled to phonons  that
optical phonon modes at the $K$ and $M$ points of the Brillouin
zone, which involves only $O$-ions displacement around a $Co$-ion,
are good candidates to display anomalies due to the CO proximity.
If by increasing of $H_2O$ content the system is pushed closer to CO,
the mentioned phonon modes
should show softening and broadening.
\end{abstract}


\maketitle

The discovery of superconductivity in
$Na_xCoO_2.yH_2O$ for $x=0.35$ and $y=1.3$ (Ref.[\onlinecite{takada03}])
has motivated a large interest in the solid state physics community due,
in part, to the similarities between this compound and high-$T_c$ cuprates.
Cobaltates are $3d$-electron systems having a quasi-two-dimensional
structure, $CoO_2$,  where
the $Co$-atoms are in a triangular lattice.
Recently, 
for describing superconductivity 
in these materials,
some authors 
proposed 
the interplay between electronic correlations 
and the proximity  
to a charge density wave (CDW) instability
\cite{motrunich04,foussats05,foussats06}. 
These papers  focus on
a two-dimensional $t-V$ model on a triangular lattice
where the nearest neighbors
Coulomb interaction $V$ 
is the relevant parameter for bringing the system to the
CDW instability which occurs for $V>V_c$ at the momentum
${\bf q}=(4\pi/3,0)$.
This CDW phase is
called $\sqrt{3}\times\sqrt{3}$-CDW
\cite{motrunich04,foussats05}.
Although 
some experimental reports in hydrated 
\cite{lemmens05,  
shimojima05} and unhydrated samples
\cite{hwang05,wu06,
qian06}
suggest a possible interpretation in favor of the CO proximity, this scenario  
is still controversial\cite{mukha}.
Therefore, it is relevant to solve the CO controversy. In this paper we propose 
that if the system is close to the CO,  
signatures of it
in the phonon subsystem can be expected.
Hence, theoretical and experimental studies on phonon
anomalies may be important for testing the CO hypothesis.
Those phonons which couple more directly with
charge fluctuation are candidates to show anomalies and are discussed in this paper.

\begin{figure}
\begin{center}
\setlength{\unitlength}{1cm}
\includegraphics[width=6.cm,angle=0.]{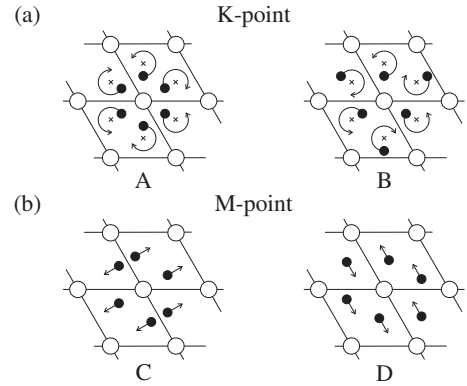}
\end{center}
\caption{a) $O$-ions pattern of displacements of the two optical
phonon modes at $K$-point discussed in the text. b) The same for the $M$-point. Big
(small full) circles represent the $Co$ ($O$) ions. The crosses
are the equilibrium positions of the $O$'s. The arrows indicate
the displacements of the $O$-ions. In the modes A and C the
average distance to the $Co$-ion of the six $O$-ions changes with
time leading to the possibility for the coupling of these modes to
the density at the $Co$ site. For B and D modes the average
distance remains almost constant. The displacements of the
$O$-ions  in the mode A have a breathing-like pattern of
displacements.
 }\label{fig1}
\end{figure}

Since the $\sqrt{3}\times\sqrt{3}$-CDW order takes place in the
$CoO_2$ planes, we have performed for simplicity a $2D$ lattice
dynamics calculation\cite{Ph} in order to understand qualitatively
which phonon modes are more sensitive to the proximity of CO. For
this purpose, we have studied the phonon
eigenvectors. For ${\bf q=0}$ we have obtained a double
degenerated optical phonon, where only the $O$-ions move,
associated with the Raman-active-in-plane  $E_{1g}$ mode
\cite{lemmens04} of frequency $\omega \sim 480 \; cm^{-1}$ and also
listed in table I of Ref.[\onlinecite{li04}]. Increasing ${\bf q}$
along this branch the degeneracy is removed. At the 
$K$-point of the Brillouin zone [${\bf q}=(4\pi/3,0)$] we have
obtained two phonon modes, A and B, with complex eigenvectors
which lead to the displacements of the $O$-ions shown in
Fig.\ref{fig1}a. At the $M$-point [${\bf q}=(\pi,\pi/\sqrt{3})$]
there are two modes, C and D, with real eigenvectors with $O$-ions
displacements presented in Fig.\ref{fig1}b. Interestingly, in the
modes A and C the average distance to the $Co$-ion of the six
$O$-ions changes with time leading to the possibility for the
coupling of these modes to the density at the $Co$ site. In
contrast, for B and D modes the average distance remains almost constant.
As it is sketched in Fig.\ref{fig1}a the displacements of the
$O$-ions  in the mode A have a breathing-like pattern of
displacements around $Co$-ions and therefore, an adequated symmetry
to couple with the charge fluctuations at the $Co$-site when the
system approaches the $\sqrt{3}\times\sqrt{3}$-CDW phase. In
addition, this mode has the same ${\bf q}=K$ vector than the CO.

For a qualitative study of those phonons which couple
more directly with density fluctuations at $Co$ site (A and C) 
we proposed the following
Hamiltonian\cite{foussats05}:

\begin{widetext}
\begin{eqnarray}
H & = & - t \sum_{<ij>,\sigma}\;( {\tilde{c}\dag}_{i\sigma}
\tilde{c}_{j \sigma}
+ h.c.) +
V \sum_{<ij>} n_i n_j
+\sum_i \omega_0 (a\dag_{i} a_i+\frac{1}{2})
+g \sum_i ({a\dag}_i+a_i)n_i
\end{eqnarray}
\end{widetext}

\noindent
where $t$ and $V$ are the hopping
and Coulomb repulsion
between the nearest-neighbors sites $i$ and $j$ on the triangular
lattice.
${\tilde{c}\dag}_{i \sigma}$ and $\tilde{c}_{i \sigma}$
are the fermionic creation and
destruction operators for holes, respectively, under the constraint
that double occupancy is excluded. $n_i$ is the fermionic density at the $Co$ site.
$a\dag_{i}$ and $a_i$ are the phonon
creation and destruction operators, respectively.
$g$ is the e-ph coupling for a given  phonon of frequency $\omega_0$.

Theoretical calculations in  Ref.[\onlinecite{roesch}] show that a
one-band $t-J$ model plus phonons can be derived, from the
three-band Hubbard model plus phonons, for studying cases where
the $O$-ions displacements lead to a modulation of the Zang-Rice
singlet energy. This view was also used in
Ref.[\onlinecite{horsch}] for studying the anomalous softening and
broadening for the half-breathing mode in cuprates.

\begin{figure}
\begin{center}
\setlength{\unitlength}{1cm}
\includegraphics[width=6cm,angle=0.]{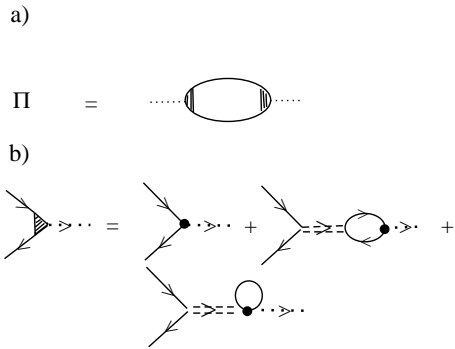}
\end{center}
\caption{a) Phonon self-energy diagram.
In $\Pi$, the bare e-ph vertex,
g (solid circle), is renormalized by
electronic correlations as showed in b). Dotted and solid lines
represent the propagators for phonons and fermions respectively.
Double dashed line is the matrix-boson propagator
\cite{foussats06} whose components
$D_{RR}$ and $D_{\lambda R}$ are associated with the charge fluctuations and the
enforcement of no-double occupancy constraint respectively.
 } \label{fig2}
\end{figure}

The phonon self-energy corrections $\Pi({\bf q},i\omega_n)$ are computed after
evaluating
the diagram plotted in Fig.\ref{fig2}a.
The dark vertex in  Fig.\ref{fig2}a is the e-ph vertex renormalized by
correlations (Fig.\ref{fig2}b).
For the evaluation of the vertex we use a recently developed $1/N$-expansion
for Hubbard operators \cite{foussats04}, where in order to get a
finite theory in the $N \rightarrow \infty$ limit,
we rescaled $t$ to $t/N$, $V$ to $V/N$, and $g$ to $g/\sqrt{N}$ and the
spin projections $\sigma$ are extended from $2$ to $N$.
See
Refs.[\onlinecite{foussats05,foussats06}] for details of the method for
the present case.
Note that phonon renormalizations are $O(1)$ which are of the same order
than the free phonon propagator (Eq.(18)  of Ref.[\onlinecite{foussats06}]).

Using Feynman rules and definitions of Ref.[\onlinecite{foussats06}]
we obtain for
the phonon self-energy $\Pi({\bf q},i\omega_n)$

\begin{widetext}
\begin{eqnarray}\label{Piph}
\Pi({\bf q},i\omega_n)=\frac{\lambda_0 \omega_0}{2 N(0)} \sum_{\bf k}
\gamma^2({\bf k},{\bf k-q};i\omega_n)
\frac{[f(E_{{\bf k-q}}-\mu)-f(E_{\bf k}-\mu)]}
{E_{{\bf k-q}}-E_{\bf k}-i \omega_n}
\end{eqnarray}
\end{widetext}

\noindent where the renormalized e-ph vertex $\gamma({\bf k},{\bf k-q};i\omega_n)$ is

\begin{eqnarray}
\gamma({\bf k},{\bf k-q};i\omega_n)&=&\frac{x}{2}[ D_{RR}({\bf q};i\omega_n)
\frac{(E_{\bf k}+E_{{\bf k-q}})}{2} \nonumber \\
& &+D_{\lambda R}({\bf q};i\omega_n)]
\end{eqnarray}

In (\ref{Piph}) $\lambda_0=2g^2N(0)/\omega_0$ is the bare
dimensionless e-ph coupling and $N(0)$ is the bare electronic
density of states (DOS). $E_{{\bf k}}= -tx(\cos k_x+2 \cos
\frac{\sqrt{3}}{2}k_y \cos \frac{1}{2}k_x)$ is the renormalized
electronic dispersion where $x$ is the electron doping. In the
following $t$ is considered to be $1$ and energies are in units of
$t$. We will present results for the commensurate doping $x=1/3$
close to the value where the highest $T_c$ is obtained in cobaltates. For
$x=1/3$ we obtain, in agreement with ARPES
experiments\cite{hasan04,yang04}, a large Fermi surface (FS)
enclosing the $\Gamma$ point. In a recent ARPES
experiment\cite{qian06}, this FS topology was relevant for the
interpretation in terms of the proximity of the system to  the
$\sqrt{3}\times\sqrt{3}$-CDW order. From $E_{{\bf k}}$ we obtain
for the dressed electronic DOS $N^*(0) \sim 0.63 1/t$ while the
bare DOS is $N(0)=x/2 N^*(0) \sim 0.1 1/t$. For the present
parameters the instability to the $\sqrt{3}\times\sqrt{3}$-CDW
order takes place for $V>V_c=1.1$ \cite{foussats05}.

\begin{figure}
\begin{center}
\setlength{\unitlength}{1cm}
\includegraphics[width=6.cm,angle=0.]{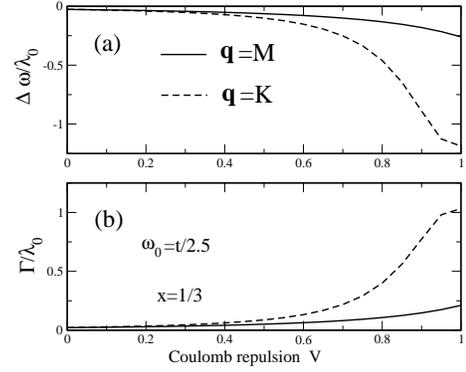}
\end{center}
\caption{a) Phonon shifting $\Delta \omega/\lambda_0$ as a function of $V$ for
the phonon modes at ${\bf q}=K$ and ${\bf q}=M$. The  phonon frequency
is $\omega_0=t/2.5$.
b) Phonon broadening $\Gamma/\lambda_0$ as a function of V.
With increasing $V$ the system approaches the CO and the two phonons soften
and broaden.
 }\label{phselfqL}
\end{figure}

The phonon frequency shifting and phonon linewidth are $\Delta
\omega= Re \Pi({\bf q},\omega_0+i \eta)$ and $\Gamma=\mathrm{Im}
\Pi({\bf q},\omega_0+i \eta)$ respectively. In Fig.\ref{phselfqL}
we show $\Delta \omega/\lambda_0$ and $\Gamma/\lambda_0$ as a
function of $V$ for the phonons at ${\bf q}=K$  and ${\bf
q}=M$. The frequency is $\omega_0=t/2.5$; being
$t\sim 150 \; meV$, $\omega_0 \sim 480 \; cm^{-1}$ which is close to the
frequency of the two optical A and C phonon modes. With increasing
$V$, approaching the $\sqrt{3}\times\sqrt{3}$-CDW phase, the two
phonons become soft (Fig.\ref{phselfqL}a) and broad
(Fig.\ref{phselfqL}b). Since the $\sqrt{3}\times\sqrt{3}$-CDW
instability occurs at ${\bf q}=K$, softening and broadening of the
${\bf q}=K$ phonon are larger than for the ${\bf q}=M$ mode.

\begin{figure}
\begin{center}
\setlength{\unitlength}{1cm}
\includegraphics[width=6cm,angle=0.]{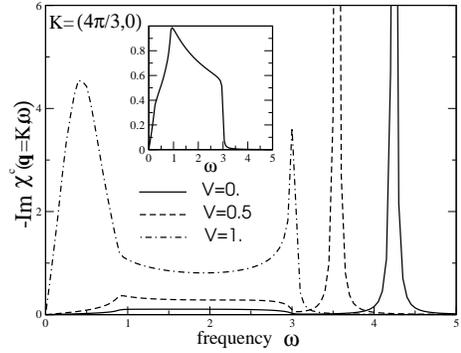}
\end{center}
\caption{Imaginary part of the density response $\chi^c({\bf q}=K,\omega)$
for $x=1/3$, at
$K=(4\pi/3,0)$, for different V, approaching $V_c$ showing the redistribution of
the spectral weight. For $V$ near $V_c$, low energy charge fluctuations renormalize
the e-ph
interaction and cause the phonon renormalizations shown in Fig.\ref{phselfqL}.
Inset: particle-hole continuum.
 }\label{sus33}
\end{figure}

The origin of the softening and broadening near the charge instability
is understood as follows.
The role played by charge densities is clearly evident by looking at eq.(3):
As stated in Refs.[\onlinecite{foussats05,foussats06}] $D_{RR}$ is proportional
to the charge-charge correlation function
$\chi^c({\bf q},\omega)=-N(\frac{x}{2})^2 D_{RR}({\bf q},\omega)$ showing that charge
fluctuations have a direct influence on the renormalized e-ph interaction.
In
Fig.\ref{sus33} are presented results for the density response, $Im
\chi^c({\bf q}=K,\omega)$, for different $V$.
The momentum was fixed at
$K$ where the
$\sqrt{3} \times \sqrt{3}$-CDW phase takes place.
For $V=0$, it is clearly seen  a
collective peak at $\omega\sim 4.2t$ at the top of the
particle-hole continuum.
With increasing $V$ the collective peak softens and for
$V=1$
is still well defined at high energy of the order
$\sim 3t$. But it has transferred spectral weight to a broad
structure at low energies with a maximum at a scale $\sim t/2$.
When the instability takes place, at $V=V_c=1.1$, this broad
structure reaches $\omega=0$ and
the static charge susceptibility diverges.
As result of this charge dynamics,
the e-ph vertex is renormalized leading to the effects shown
in Fig.\ref{phselfqL}. 
Since the instability is of static character, the effect
will be stronger for low energy phonons than for large energy phonons.
Contrary to the case for 
quarter filling on the square
lattice\cite{merino03}, 
the charge instability at
$x\sim 1/3$ can not be seen strictly
as the softening of a
collective charge mode, but a redistribution of spectral weight
takes place.
The reason for this behavior can be found
in the form of the particle-hole
continuum
which starts
at $\omega=0$ (inset in Fig.\ref{sus33}) and then,  the collective
peak never merges from the bottom
of the continuum.
Finally, it is important to notice that
the static CO was not experimentally confirmed (see Ref.\onlinecite{mukha}), 
however   
the effects discussed in present paper  
need only a dynamical 
CO proximity in which the electronic system remains homogeneous.  

To our knowledge the exact value of the e-ph coupling is not known
for cobaltates, however recent experiments suggest that this is
nonegligeable\cite{lupi04}. In
Refs.[\onlinecite{foussats05,foussats06}], for estimations of
$T_c$, we have proposed for the total e-ph coupling 
$\lambda=0.4$ which is close to the order
of other recent estimates\cite{yada05,rueff06}. In the context of
superconductivity, this value must be related with  
$\lambda=\sum_{\nu} \lambda_{\nu}$ where $\nu$ runs over the
number of phonon branches and $\lambda_{\nu}$ is the coupling for
each branch. Considering for instance $\lambda_0=0.04$, which is
one tenth of $\lambda$, we have found, for $V \sim 0.9$, a phonon
frequency shifting $\Delta \omega \sim 50 \; cm^{-1}$ ($15 \; cm^{-1}$)
for the $K$ ($M$) phonon with respect to the situation for
$V<<V_c$. A similar estimate can be
done for $\Gamma$ showing a  broadening of the order of $\Gamma
\sim 25 \; cm^{-1}$ ($6  \; cm^{-1}$) for the $K$ ($M$) phonon mode with
respect to weakly or unhydrated samples. In spite of the crudeness
of the estimate, it shows that the effect is large enough as to be
observed.

It is important to make one remark. For ${\bf q}=0$, in the $E_{1g}$ mode
$O$-ions move parallel to the planes compressing the $Co$-atoms.
However, from eq.(2) we expect phonon anomalies to be
weaker for ${\bf q} \sim 0$ modes than for large ${\bf q}$ ones.
Therefore, if ${\bf q}\sim0$ phonons
probe the charge fluctuations due to the proximity to CO, it is through
inter-band
transitions while, for large ${\bf q}$ phonons, intra-band transitions
are directly involved
and we expect the effect to be larger.

Let us discuss how the presented results may be 
verified experimentally. 
When studing superconductivity\cite{foussats05,foussats06}
we obtain an increment of $T_c$ when $V$ increases to $V_c$. 
Based on the experimental results showing that 
$T_c$ increases with the
water content $y$
\cite{sakurai04}  
we 
suggest 
a positive
correlation between $V$ and $y$.
On the other hand, and in spite of  
a microscopic 
relation between $V$ and $y$ is lacking,
experiments
\cite{sakurai04} and theory\cite{marinetti04} 
show that hydratation causes the electronic structure to 
become more two dimensional and then, one may argue that the Coulomb interaction 
between   
nearest-neighbors 
is less screened. 
Then, if hydration pushes the system closer to CO, we expect 
larger phonon anomalies for samples with large $y$.
Since the 
predictions in present paper are for large
${\bf q}$ optical
phonons, they are in general accessible to
neutron scattering experiments and, more recently, 
to inelastic x-rays scattering
\cite{rueff06}. 
Since hydratation expands the lattice, mainly in the 
$c$-direction\cite{sakurai04}, one could expect 
that phonon dispersions also change due to this effect, making hard  
the detection of the predictions of the present paper. 
However, as the phonons modes discussed here are mainly of two dimensional 
character and perpendicular to the $c$-axis, we expect our result to be dominant.

In summary, we have studied the influence of
charge fluctuations, near the CO instability, on optical phonons
modes at $K$ and $M$ points of Brillouin zone.
We have discussed two in-plane optical
phonons  where only $O$-ions move compressing the $Co$-atoms.
These phonons were predicted to be
soft and broad when the systems is pushed near the charge instability.
Softening and broadening were found to be larger for $K$ than for $M$ phonon
because the former has the correct ${\bf q}$ and symmetry to couple with the
$\sqrt{3}\times\sqrt{3}$-CDW phase.

The authors thank to M. Cardona, P. Lemmens, H. Parent, 
and M. Stachiotti for valuable discussions. We specially thank to A. Muramatsu for
illuminating discussions.


\begin{thebibliography}{10}
\bibitem{Ph}
We have considered a $2D$ $CoO_2$ plane where the $Co$-atoms are
on a triangular lattice and $O$'s in the center of each triangle. 
$Co$ and $O$ masses were
chosen $m_{Co}=1$ and $m_O=0.27$ respectively. We have also
considered longitudinal ($K_L$) and transversal ($K_T$) force
constants between nearest-neighbors $Co-O$ and $O-O$:
$K_L^{Co-O}=1$, $K_T^{Co-O}=0.5\; K_L^{Co-O}$, $K_L^{O-O}=0.1\;
K_L^{Co-O}$ and $K_T^{O-O}=0.1 \;K_T^{Co-O}$.
\end{thebibliography}
\end{document}